\documentclass[multphys,vecphys]{svmult}

\usepackage{makeidx}         
\usepackage{graphicx}        
\usepackage{multicol}        
\usepackage[bottom]{footmisc}

\makeindex             

\begin{document}

\title*{Star formation histories of resolved galaxies}


\author{Monica Tosi
}


\institute{INAF - Osservatorio Astronomico di Bologna, Via Ranzani 1, Bologna,
Italy
\texttt{monica.tosi@oabo.inaf.it}
}
%
%
\maketitle

\begin{abstract}
The impact of HST photometry and European astronomy in studies concerning
the star formation histories of resolved galaxies is described.  
Our current knowledge of the
star formation history of systems within 10-20 Mpc, as derived from the
colour-magnitude diagrams of their resolved stellar populations, is reviewed,
as well as the impact of these results on our understanding of galaxy
evolution.

\end{abstract}

\section{Introduction}
\label{sec:1}

One of the astronomy research fields which have been more impressively impacted 
by the advent of HST is the study of the star formation histories (SFH) of 
resolved stellar populations. The superior photometric performances of HST have
allowed the resolution of individual stars to unprecedented (and still
unequalled) levels of faintness, distance and crowding. Such an outstanding 
resolving power has triggered a large number of HST programs aimed at the 
derivation of the SFH of galaxies in
the local Universe, involving many people on both sides of the Atlantic
Ocean. These studies have significantly improved our understanding of galaxy
formation and evolution. The impacts of European
Astronomy and of HST in the field are briefly summarized in the following.
 
\section{The impact of European Astronomy}
\label{sec:2}

\begin{figure}
\centering
\includegraphics[height=13cm]{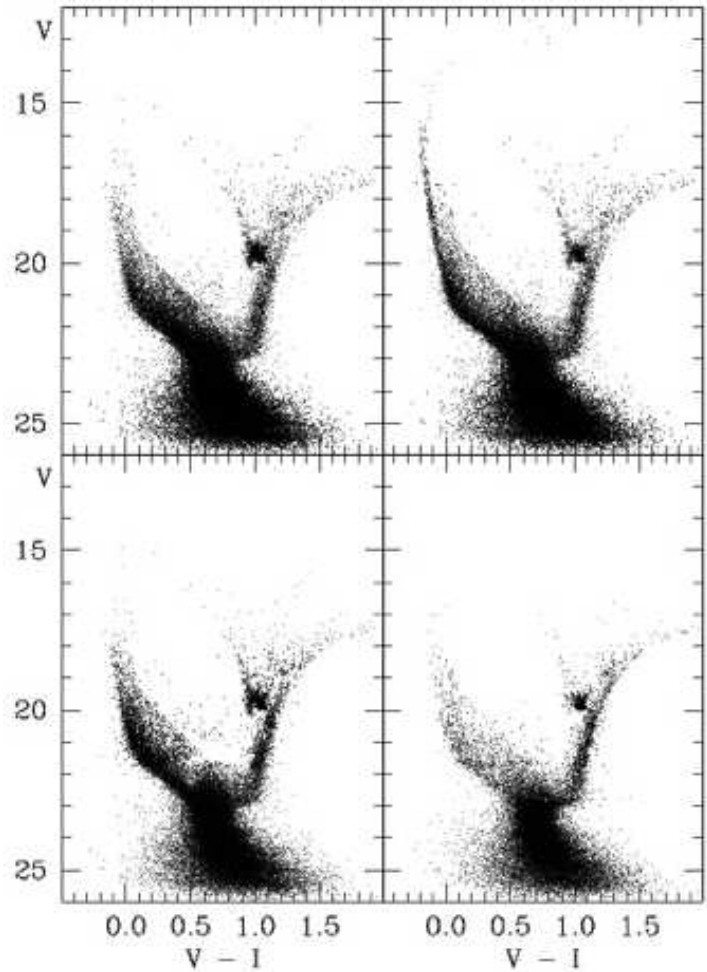}
\caption{The effect of the SFH on the CMD of a 
hypothetical galactic region 
with (m-M)$_0$=19, E(B-V)=0.08, and with the photometric errors and 
incompleteness typical of HST/WFPC2 photometry. All the shown synthetic
CMDs contain 50000 stars and are based on the Z=0.004 Padova models \cite{F94}.
Top-left panel: the case of a SFR constant from 13 Gyr ago to the present epoch.
 Top-right panel: the effect of concentrating in the last
20 Myr the formation of all the stars which were born in the last 100 Myr in
the previous case. The CMD has a much brighter and thicker blue plume.
Bottom-left panel: same constant SFR as in the
first case, but with a quiescent interval between 3 and 2 Gyrs ago; a gap appears
in the CMD region delimited by the 2 and 3 Gyr isochrones. Bottom-right panel:
exponentially decreasing SFR over the whole Hubble time, with e-folding time 5
Gyr; the blue plume contains much fewer stars than in the constant SFR case, and
the late evolutionary phases (RGB, red clump, blue loops) are also differently 
populated, with fewer stars in the younger sequences. }
\label{syn1}       
\end{figure}

The whole business of deriving the SFH of nearby galaxies from the 
colour-magnitude diagrams (CMDs) of their resolved stars started in Europe, 
when the synthetic CMD method was developped several years ago 
\cite{T91, Be92, Ga96, T96}. 
The method is based on the well known circumstance that the location of any 
star in the CMD is uniquely related to its mass, age and chemical composition.
This makes the CMD the best tool to infer the age of stellar systems. In the
case of simple stellar populations, i.e. coeval stars with the same chemical
composition, isochrone fitting is the most frequently used method to infer the
system age. In the case of galaxies, with rather complicated mixtures of
different stellar generations, the age determination is less straightforward,
but their CMDs remain the best tool to derive the SFH.  

Fig.\ref{syn1} shows the
effect of different SFHs on the CMD of a hypothetical galactic region with the
intrinsic and photometric 
properties typical of a region in the SMC observed with
HST/WFPC2. If the star formation rate (SFR) has been constant for all the galaxy
lifetime, the CMD of the region is expected to have the morphology of the
top-left panel, with a prominent blue plume mostly populated by main-sequence
(MS) stars and an equally prominent red plume resulting from the overposition
of old subgiants, red giant branch (RGB), horizontal branch (HB) and asymptotic giant 
branch (AGB) stars, with younger stars in the AGB, blue loops and red
supergiant phases. If we change the SFH, not dramatically, but simply introducing
quiescent and/or more intense phases, or assuming a shorter e-folding time for 
the SFR, the morphology of the CMD changes significantly, as visible in the 
other panels of Fig.\ref{syn1}. The
tight dependence of the CMD morphology on the SFH is the cornerstone of the
synthetic CMD method, which consists in comparing the observational CMD of a
galactic region with synthetic CMDs created via MonteCarlo extractions on
stellar evolution tracks or isochrones for a variety of SFHs, IMFs and
age-metallicity relations (see e.g. \cite{T91, G98, ag04} for a detailed
description of different procedures). 

The method does not provide unique solutions, but significantly reduces 
the possible SFH scenarios. Since its first applications to photometric data 
from ground-based, moderate size telescopes, the synthetic CMD method
demonstrated its power, showing that even in tiny galaxies such as Local Group 
dwarf
irregulars (dIrrs) the SFH varies from one region to the other and that their
star formation regime is rather continuous, with long episodes of moderate
activity, separated by short quiescent intervals (the so-called gasping regime,
 \cite{F89, T91, M95, Ga96, T96}).

\section{The impact of HST}
\label{sect:3}

When the first non-aberrated images were acquired with HST, the impressive
improvement in the achievable photometric resolution and depth, and the
corresponding quantum leap in the quality of the CMDs, triggered a worldwide
burst of interest in the derivation of the SFHs of nearby galaxies and in the
synthetic CMD method. Many people developped their own procedures and a few
years later already $\sim$10 different groups participated to an experiment
organized in Coimbra (Portugal) to compare with each other the results from 
different procedures (see \cite{SG02} and references therein). The experiment
showed that most of the procedures provided consistent results and brought
further interest to the field. Since then, many studies have been performed to
infer the SFH of almost all the dwarf galaxies in the Local Group (e.g. 
\cite{Do03, Sk03, HZ04, Mom05, Co07}).

The high spatial resolution of the HST cameras also allows us to 
spatially resolve the SFH  in the closest 
galaxies. For instance \cite{DP98} and \cite{DP02} have  resolved
and measured the SF activity over the last 0.5 Gyr in the various sub-regions 
of the dIrrs Gr8 and Sextans A, close to the borders of the Local Group.  
The resulting space/time distribution of the SF, with lightening and fading of 
adjacent cells, is
intriguingly reminiscent of the predictions of the stochastic self-propagating
SF theory proposed by \cite{SSG79} almost 30 years ago.
 
The HST/ACS provides spectacularly deep and spatially resolved images, such
as those of the star forming region NGC346 in the SMC, where \cite{No06} 
and \cite{S07} have been able to measure 85000 stars, 
from very old to very young ones, including, for the first time in the SMC,
pre-main-sequence objects of mass from 3 to 0.6 M$_{\odot}$. The famous CMD of
an Andromeda region by \cite{Br03} showed that detailed SFHs can be finally 
derived also for external spirals.

In galaxies beyond the Local Group, distance makes crowding more severe, 
and even 
HST cannot resolve stars as faint as the  MS turn-off  of old
populations. The higher the distance, the worse the crowding conditions, and the
shorter the lookback time reachable even with the deepest, highest resolution
photometry (see Fig.2 in \cite{To07}).  Depending on distance and intrinsic crowding, the reachable 
lookback time in galaxies more than 1 Mpc away ranges from several 
Gyrs (in the best cases, when the RGB or even the HB
 are clearly identified), to several hundreds Myr (when AGB stars are 
 recognized), to a few tens Myr (when only the brightest
supergiants are resolved). 

   \begin{figure}
   \centering
   \includegraphics[width=10cm,height=7.8cm,clip]{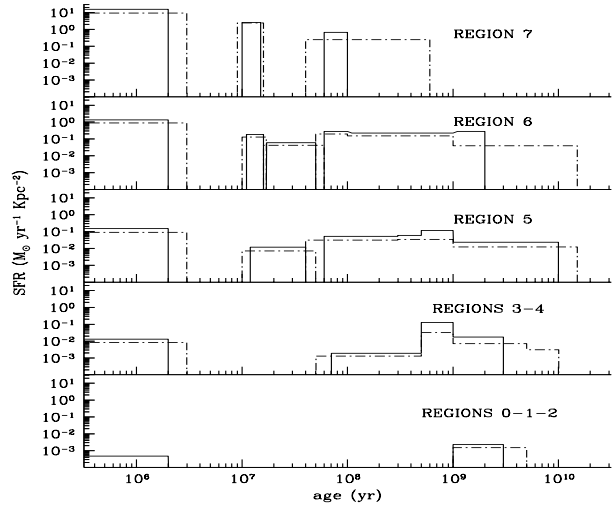}
     \caption{SFR per unit area vs age for NGC1705's concentric regions
     (decreasing numbers from the center outwards). 
     Notice that both axes are on
     logarithmic scale. The different linetypes 
     show the uncertainty on the inferred SFH \cite{An03}.} 
    \label{sf1705}
    \end{figure}
Since the Local Group doesn't host all types of galaxies, with the notable and
unfortunate absence of the most (ellipticals) and the least (Blue Compact
Dwarfs, BCDs) evolved ones, a few groups have embarqued in the challenging 
task of deriving the SFH of more distant galaxies. In spite of the larger 
uncertainties and the shorter lookback time, these studies have led to quite 
interesting results, which wouldn't have been possible without HST.

The case of NGC1705 is particularly instructive. The HST/WFPC2
photometry was deep and good enough to let us resolve individual stars from 
the most central regions to the extreme outskirts \cite{To01}. 
We have been able to divide the galaxy in 
roughly concentric regions, all sufficiently populated by resolved stars,
and derive the SF history of each region. Young massive stars are concentrated
at the center and their percentage rapidly decreases outwards, while faint
red stars are increasingly visible towards the outer regions. The latter
circumstance doesn't necessarily imply that old stars are absent at the center;
it simply means that crowding is too severe there to let us resolve them.
In the outer regions, where crowding is definitely not a problem, the CMDs
present a well defined upper portion of the RGB, whose
tip is also very well defined and allowed us to accurately derive the galaxy 
distance \cite{To01}.

By applying to each region the synthetic CMD method, \cite{An03}
have inferred their SFHs, summarized in Fig.\ref{sf1705}, where the SFR
per unit area is plotted as a function of age. It can be seen that, except
for the innermost region, where crowding does not allow us to reach old lookback
times, all the regions have been forming stars since at least 5 Gyr.
On average, the SF appears to have been rather continous:
there are evidences for interruptions in the SF activity, but always shorter
than a few Myr or tens of Myr, at least in the age range where we do have this
time resolution (i.e. in the last 1 Gyr or so). Quiescent phases of 100 Myr or
longer would have appeared as gaps in the empirical CMDs of stars younger than
1 Gyr, and such gaps are absent. The SF history of NGC1705 shows three striking
features: one is the burst occurred in the central regions 10--15 Myr ago, when
the central Super Star Cluster also formed and when the observed galactic wind 
is supposed to have
originated; the second is the quiescent phase with no SF anywhere in the galaxy
right after such burst, probably due to the gas sweeping by shocks and winds 
triggered by
the explosions of the burst supernovae; and the third is
the new, even stronger, SF activity occurring everywhere in NGC1705 (but much
higher in the inner regions) in the last 2 Myr. The latter event puts
interesting constraints on the cooling timescales of the gas heated by the
supernovae generated in the previous burst and on the modeling of SF processes.

\begin{figure}
\centering
\includegraphics[height=6.cm]{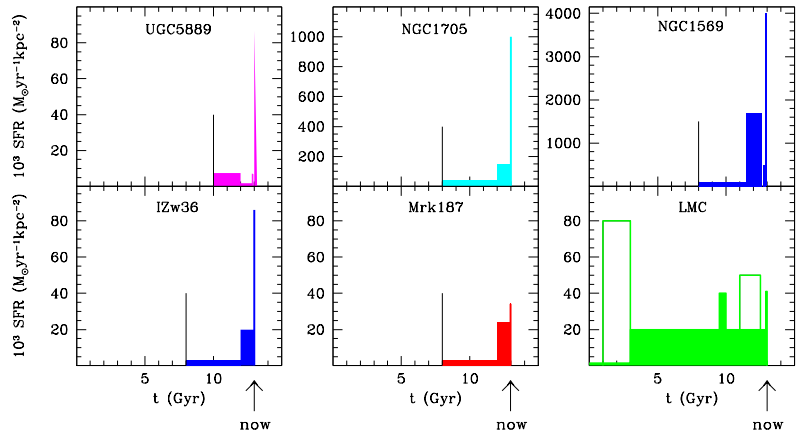}
\caption{SFH (SF rate per unit area vs time) from the CMDs of a few late-type
galaxies resolved by HST: three BCDs (NGC~1705 \cite{An03}, IZw36 \cite{Sl01},
Mrk187 \cite{Sl00}), one starburst dIrr (NGC~1569 \cite{G98, An05}), 
one LSB (UGC5889 \cite{Va05}), and the LMC bar used for the Coimbra experiment 
\cite{To02}. The lookback time reached by the photometry is indicated in each
case with a black vertical line. Notice that these are global SFHs (except for
the LMC bar), while Fig.2 shows the SFH of individual regions.}
\label{sfall}       
\end{figure}

Fig.\ref{sfall} sketches the global SFHs derived by various authors for some 
of the late-type dwarfs studied so far. The lookback time is indicated and 
in all cases stars with that age were detected.

The latter is one of the most interesting results of all the studies of BCDs 
whose individual 
stars have been resolved by HST and the SFH has been derived with the 
synthetic CMD method: all of them have
turned out to be already active at the lookback time reached by the photometry
(see \cite{L98, Al99, C00, Sl00, D01, Sl01, 
C02, D02, An03, Al05}). None appears to be 
experiencing now its first star formation activity, including the most 
metal poor, gas rich ones, such as SBS1415 and
IZw18 (see \cite{Al05, Mom05, To06} and Aloisi, this volume). 

Equally interesting is that the galaxy with
stronger SF activity is the dIrr NGC1569, the only one with a rate
comparable to the 1 $M_{\odot}yr^{-1}$ required by \cite{BF96}
to let a late-type galaxy contribute to the blue galaxy excess observed
in counts at intermediate redshift. All the other dwarfs, independently of being
classified as BCDs or irregulars or low surface brightness (LSB), present much 
lower SFRs. 

We also notice that all the galaxies of Fig.\ref{sfall}, but the LMC, have 
similar global SFHs, with a recent burst overimposed on a quiter, more
continuous regime. If we consider also the other late-type dwarfs in the Local
Group studied so far, we can speculate that the only apparent
difference in the SFH of BCDs and dIrrs is that the former always show a very 
recent SF burst, the latter not always. This is presumably due to  
selection effects that made it possible in the past to discover distant dwarfs
only if featured with strong HII region emissions, whilst nearby dwarfs
were discoverable also if the peak of their SF activity occurred long ago.

\section{The impact on galaxy evolution}

The main results from all the studies performed on the SF histories 
of dwarf galaxies are: 1)
 no evidence of long interruptions in the SF activity has been found,
except in early-type galaxies;
2) no galaxy currently experiencing its first SF activity has been found yet;
3) strong bursts don't seem to be frequent;
4) the SF regime seems to be a gasping rather than a bursting one in all   
late-type dwarfs, both in the Local Group and outside it;
5) no significant difference has been found in the SFH and in the
stellar populations of BCDs and dIrrs, except that the former always have a
recent SF burst.

These results have profound effects on our understanding of the evolution of
dwarf galaxies. In addition, the combination of HST photometry with 
high-performance
spectroscopy  is providing in the last years
a first in-depth view of their chemical evolution. The detailed chemical 
abundances measured in many of their stars have solved  the
age-metallicity degeneracy affecting the RGB colours and have led to the
derivation of age-metallicity relations not only in the LMC (e.g. \cite{Co05})
but also in other nearby dwarf spheroidals (e.g. \cite{To03}).
As a consequence, for the first time detailed chemical evolution models can be
computed for individual dwarf galaxies, an application performable until 
recently only for spirals, where the wealth and quality of the observational
data were sufficient to properly constrain the models. The new
generation models can now adopt SF laws and initial mass functions which are not
free parameters, but are provided by the SFH studies, and their predictions 
can be compared with
the chemical abundances not only of very young objects, but also of stars of
various ages. Such a decrease in the number of free
parameters opens the route to a much more reliable use of theoretical models
and, ultimately, to a much better understanding of galaxy evolution.

\bigskip\noindent
I'm grateful to all the persons I have studied the SFHs with, in 
particular to  A. Aloisi, L. Angeretti, F. Annibali, L. 
Greggio, A. Nota and E. Sabbi. This work has been partially supported by the 
PRIN-INAF-2005.

%
%
%



\printindex
\end{document}